\documentclass[12pt]{article}
\usepackage{latexsym}

\begin{document}

\newcommand{\be}{\begin{equation}}
\newcommand{\ee}{\end{equation}}
\newcommand{\<}{\langle}
\renewcommand{\>}{\rangle}
\newcommand{\reff}[1]{(\ref{#1})}

\title{The transfer matrix with Kogut-Susskind fermions} 
\author{
  { Fabrizio Palumbo~\thanks{Work supported in part by the 
European Community's Human Potential Programme under contract
HPRN-CT-2000-00131 Quantum Spacetime}}            
\\[-0.2cm]
  {\small\it INFN -- Laboratori Nazionali di Frascati}  \\[-0.2cm]
  {\small\it P.~O.~Box 13, I-00044 Frascati, ITALIA}          \\[-0.2cm]
  {\small Internet: {\tt palumbof@lnf.infn.it}}     
   }
\maketitle

\thispagestyle{empty}   


\begin{abstract}
A transfer matrix for gauge theories on a lattice with Kogut-Susskind fermions in the 
flavour basis is constructed and some difficulties in the spin-diagonal basis are
pointed out.
\end{abstract}

\clearpage

\section{Introduction}

In the investigation of some features of QCD at finite baryonic density~\cite{Palu} we
must be able to write the baryon charge operator in the transfer matrix formalism.
Therefore we must express the transfer matrix in terms of quark-antiquark creation-annihilation
operators. Such a formulation has been constructed by by L\"uscher~\cite{Lusc} for Wilson
fermions, but we have not found in the literature anything similar for Kogut-Susskind fermions.
Actually we know only one paper on the subject~\cite{Shar}, but in this work the transfer matrix
is constructed in the spin-diagonal basis, where it is found necessary to introduce auxiliary 
fields whose relation to quark-antiquarks is not transparent. Another feature, possibly related
to the above, might be a difficulty in the way of the applications we are interested in: the
transfer matrix is linear in the creation-annihilation operators of quark-antiquarks, rather
than exponential, so that the hamiltonian  contains the logarithm of these operators.

We have therefore tackled the problem in the flavour basis, and we have extended the 
formalism developed by L\"uscher for Wilson fermions to the
case of Kogut-Susskind under the condition that the elementary gauge variables be
attached to the 
links of the blocks rather than  of the lattice. Let us explain the reason of this
restriction. In the Conclusion we will point out its consequence.

 L\"uscher prooved the reflection positivity~\cite{Lusc} of the transfer matrix 
in the gauge $U_0 = 1\!\!1$. 
This gauge fixing is too strong, but Menotti and Pelissetto~\cite{Meno} have extended the 
proof to the  gauge $U_0 \sim 1\!\!1$, where $U_0 = 1\!\!1$ with
the exception of a single time slice. If for Kogut-Susskind fermions we define the gauge
fields on the links of the blocks, nothing changes in the Menotti-Pelissetto' proof 
which we have used in our construction. But if we
associate the gauge fields to the links of the lattice, their gauge fixing becomes 
$U_0(n_0)U_0(n_0+1) \sim 1\!\!1$ (in standard notation which is however spelled out below). 
This has a nontrivial Faddeev-Popov determinant which  has not been evaluated but must 
be taken into account to establish the reflection positivity of the transfer matrix.

In this paper we will ignore the pure gauge field part of the action, the partition
function and the transfer matrix. Its role in fact is the same as with Wilson fermions,
and it can be included in the  same way as in L\"uscher's paper.

\section{The transfer matrix}

The quark field $q$ in the flavour basis is defined on hypecubic blocks of coordinates
$n_{\mu}$ defined in terms of the sites coordinates $x_{\mu}$
\be
x_{\mu}= 2 n_{\mu}+
\eta_{\mu},\,\,\,x_{\mu}=0,...,2 N_{\mu}-1,\,\,\,n_{\mu}=0,...,N_{\mu}-1,\,\,\,
\eta_{\mu}=0,1.
\ee
It carries Dirac indices $\alpha$, flavour indices $f$ and color indices $c$, sometimes
comprehensively denoted in the sequel by $I$. 

 In the transfer matrix formalism most often one has to do with quantities at
a given (euclidean) time  $n_0$. For this reason we use a summation convention
over spatial coordinates ${\bf n}_j$ and intrinsic indices $I$ at fixed time. 
So for instance we will write
\be
\overline{q}(m_0) Q(m_0,n_0) q(n_0)= \sum_{{\bf m},{\bf n},I,J} 
\overline q_{{\bf m},I}(m_0)
Q_{{\bf m},I;{\bf n},J}(m_0,n_0) q_{{\bf n},J}(n_0),
\ee
where $Q$ is the quark matrix. It is a function of
time dependent link operators $U_{\mu}(n_0)$ which have the  standard Wilson variables $
U_{\mu}(m_0,{\bf m}) $ as spatial matrix elements
\be
\left(U_{\mu}(m_0) \right)_{{\bf m},{\bf n}} = \delta_{{\bf m},{\bf n}} U_{\mu}(m_0,{\bf m}).
\ee

In this notation the quark action can be written
\be
S_q = \sum_{m_0,n_0}\overline{q}(m_0) Q(m_0,n_0) q(n_0). 
\ee
Its explicit expression~\cite{Mont} is
\be
S_q=  \overline{q} \left\{ M 1\!\!1\otimes 1\!\!1 + 2 K \left[ \gamma_{\mu} \otimes 1\!\!1
\nabla_{\mu} - \gamma_5 \otimes t_5 t_{\mu} \delta_{\mu} \right] \right\} q,
\ee
where the lattice spacing is equal to 1, $M$ and $K$ are the mass and the hopping parameters
and $\nabla_{\mu}$ and $\delta_{\mu}$ are the derivatives
\begin{eqnarray}
 \left( \nabla_{\mu} f \right) (n) &=&  { 1 \over 4}\left( f(n+2\mu)-f(n-2\mu) \right) 
\nonumber\\
\left(\delta_{\mu}f \right)(n)  &= & { 1 \over 4}\left( f(n+2\mu)+f(n-2\mu) -2f(n) \right). 
\end{eqnarray}
 In the tensor product the
$\gamma$-matrices act on Dirac indices, while the $t$-matrices act on flavor indices and are
given by
\be
t_{\mu}= \gamma_{\mu}^T.
\ee
The minimal gauge interaction (that where the link variables join the quark fields along the
shortest path) can be introduced in the standard way by means of the block link variables
$U_{\mu}$. 

Now we want to identify the components of the quark field which can be related to
quarks and antiquarks. They propagate respectively
forwards and bakwards in time and are interchanged by charge conjugation. For this purpose we
define the projection operators
\be
P_{\mu}^{(\pm)}={ 1\over 2} \left[ 1\!\!1 \otimes 1\!\!1 \pm
\gamma_{\mu} \gamma_5 \otimes  t_5t_{\mu} \right].
\ee
 We can thus rewrite the quark matrix singling out the terms which
are not diagonal in time in the form
 \begin{eqnarray}
Q(m_0,n_0) &=&  K \left[ \gamma_0 P_0^{(-)} U_0(m_0) \delta_{n_0,m_0+1}  - \gamma_0 P_0^{(+)}
  U_0^{(+)}(n_0) \delta_{n_0,m_0-1} \right]
\nonumber\\
 & & +  N(n_0) \delta_{m_0,n_0},
\end{eqnarray}
where
\begin{eqnarray}
N(n_0) &= &  M  \, 1\!\!1 \otimes 1\!\!1 + K \gamma_5 \otimes t_5 t_0 
+ K \sum_{j=1}^3 \left\{   \gamma_5 \otimes t_5 t_j  \right.
\nonumber\\
 & & \left. +  \gamma_j \left[  P^{(-)}_j U_j(n_0) T^{(+)}_j -P^{(+)}_j T^{(-)}_j U_j^+(n_0)
\right] \right\}.
\end{eqnarray}
The translation operators $T_j^{(\pm)}$ have matrix elements
\be
\left( T_j^{(\pm)} \right)_{{\bf m},{\bf n}} =  \delta_{{\bf n},  {\bf m} \pm  e_j}
\ee
with
\be
\left( e_j \right)_k = \delta_{j,k}.
\ee
We see that the components 
\be
q^{(\pm)} = P_0^{(\pm)} q
\ee 
of the quark field
propagate backwards/forwards respectively. It remains to check that they are properly
related by charge conjugation. A possible definition of this operator, with L\"uscher's
convention for the
$\gamma$-matrices is
\be
{\cal C}= \gamma_0 \gamma_2 \otimes t_0t_2
\ee
whose action on the quark field is
\begin{eqnarray}
q &=& {\cal C}^{-1} \, \overline{q}\,'
\nonumber\\
 \overline{q} &=& -  \, {\cal C}^T \,q\,'.
\end{eqnarray}
Therefore charge conjugation
\be
q^{(+)}  = \gamma_2 \otimes t_0 t_2 \left(q^{(-)} \right)'
\ee
amounts to interchange $q^{(\pm)}$ with one another and to rotate the flavours. This action 
 is accettable  insofar as 
the flavour is physically unobservable.

Finally following the L\"uscher's procedure we define the transfer matrix $\hat{{\cal T}}_q$ in
terms of creation and annihilation  operators related to $q^{(\pm)}$ and show that
\be
Z_q= Tr \,  \hat{\cal{T}}_q = \int [ d \overline {q} d q] \exp (- 16 \, S_q). \label{trace}
\ee
The factor of 16 in front of $S_q$ is the volume of the block.
We construct  $\hat{{\cal T}}_q$ in terms of the auxiliary operator
 \be
 \hat{T}_q(n_0)  =   \exp\left( \hat{y}N(n_0) \hat{x} \right)
\ee
according to the ordered product
\be
\hat{\cal{T}}_q = {\cal J}\prod_{n_0=0}^{N_0-1} \left(\hat{T}_q(n_0)\right)^+
   \, \hat{T}_q(n_0 - 1). \label{transfmatr}
\ee
In the above equations
 ${\cal J}$ is the jacobian of a transformation which will be defined later, and
the operator $ \hat{ T}_q(n_0) $ depends on the time $n_0$ only throu the dependence 
on it of the gauge fields. Because $N$ is hermitean, $\hat{\cal{T}}_q$ is positive definite. 
 The above expression of $\hat{T}_q(n_0)$ is valid in the 
gauge $U_0 = 1\!\!1$, assumed 
by L\"uscher to prove reflection positivity. As already said this gauge fixing is too strong,
but to
lighten the formalism we will nevertheless put $U_0= 1\!\!1$ and we will reinstate $U_0$  at the
end. The reader can check that keeping $U_0$ in the intermediate steps one arrives at the same
result.

To prove Eq.~\reff{trace} we must transform the trace into a Berezin integral. To
this end we introduce between the factors in Eq.~\reff{transfmatr} the identity
\be
1\!\!1 = \int [dx^+ dx \, dy^+ dy] \exp(-x^+ x - y^+ y) |x \, y><x \, y|,
\ee
where the basis vectors
\be
|x \, y > = |\exp( - x \,\hat{x}^+  - y \hat{y}^+) >
\ee
are coherent states and the $x,y$ are Grassmann variables. We then
 use the following equations. First
\be
<x   |x'> = \exp \left( x^+  x' \right).
\ee
 Second, if $B=B(\hat{x}^+)$ and $C = C( \hat{x})$ are operators which depend
on $ \hat{x}^+, \hat{x}$ only
\be
<x|B \cdot C |x'>= B( x^+) <x|x'> C(x').
\ee
By means of these equations we can evaluate the kernel of the transfer matrix
\begin{eqnarray}
& &<x(n_0),y(n_0)|\hat{{\cal T}}_q(n_0) |x(n_0-1),y(n_0-1)> = {\cal J}^{-1}
 \exp\left( x^+(n_0) N(n_0) y^+(n_0) \right)
\nonumber\\
& & \,\,\,\,\,\,\,\,\,\,
 \cdot  
\exp\left( x^+(n_0)  \, x(n_0-1) +(y^+(n_0)   \, y(n_0-1) \right)
\nonumber\\
 & & \,\,\,\,\,\,\,\,\,\,  \cdot \exp\left(y(n_0-1) N(n_0-1) x(n_0-1)\right).  
\end{eqnarray}

Collecting all the pieces we arrive at the euclidean path integral form of the partition function
\be
 Z_q = {\cal J}^{-1} \int [dx^+ d x \, d y^+ d y] 
\exp  S'   \label{Seff}
\ee
with the action
\begin{eqnarray}
S' &=& \sum_{n_0} -x^+(n_0) x(n_0-1) -y^+(n_0) y(n_0-1)
+ x^+(n_0) x(n_0) +y^+(n_0) y(n_0)
\nonumber\\
& & - x^+(n_0) N(n_0) y^+(n_0) - y(n_0-1) N(n_0-1) x(n_0-1) .
\end{eqnarray}
 The change of variables
 \be
x =  4 \sqrt{K}q^{(+)},\,\,\,y^+=  4 \sqrt{K}q^{(-)},
\ee
whose jacobian is the function ${\cal J}$ introduced in Eq.\reff{transfmatr}
 transforms $S'$ into $S_q$ thus proving Eq.\reff{trace}.

\section{Conclusion}

We have extended the Lusher's formalism for Wilson fermions to the case of Kogut-Susskind 
in the flavour basis under the
condition that the gauge fields be associated to the links of the blocks. This restriction is
due to the use of the Menotti-Pelissetto proof of reflection positivity which we used in the
construction.

This result can be sufficient for our purposes~\cite{Palu} but leaves unsolved the case of the
spin-diagonal basis. In fact we cannot get the spin-diagonal formulation by a simple change 
of basis~\cite{Mont}. There is a formal difficulty and a practical complication in the way. To
this end
 one should associate the gauge fields to the  links of the lattice and this
requires an extension of the proof of reflection positivity in the presence of the Faddeev-Popov
determinant in the gauge $U_0(n_0)U_0(n_0+1) \sim 1\!\!1$. Assuming this to be done, 
however, we 
would get an awkward result because a minimal gauge coupling  in one basis becomes  cumbersome
in the other~\cite{Klub}. One should then start from the minimal gauge coupling in the
spin-diagonal basis, find the coupling resulting in the flavour basis, and show reflection
positivity with such coupling. Both shortcomings are hopefully  surmountable, but the problem
requires further study  and we think that deserves a wider interest than our specific 
motivation.

The alternative is to construct the transfer matrix directly in the spin-diagonal basis
in a form convenient for the use at finite baryon density.


\clearpage 

\begin{thebibliography}{9}

\bibitem{Palu}
F.Palumbo, hep-lat/0202021; hep-lat/0208002

\bibitem{Lusc}
M.L\"uscher, Commun. Math. Phys. 54 (1977) 283

\bibitem{Shar}
H.S.Sharatchandra, H.T.Thun and P.Weisz, Nucl.Phys.B192(1981)205

\bibitem{Meno}
P.Menotti and A.Pelissetto, Commun. Math. Phys.113(1987)369

\bibitem{Mont}
I.Montvay and G.M\"unster, Quantum fields on a lattice, Cambridge University press, 1994

\bibitem{Klub}
H.Kluberg-Stern, A.Morel, O.Napoly and B.Petersson, Nucl. Phys. B220[FS8] (1983) 447;
G.T.Bodwin and E.V.Kovacs, Phys. Rev. 38D (1988) 1206


\end{thebibliography}
\end{document}